\documentclass[floatfix,showpacs,preprintnumbers,amsmath,amssymb,aps,twocolumn,superscriptaddress]{revtex4}

\usepackage{amsfonts}
\usepackage[pdftex]{graphicx}
\usepackage{color}
\usepackage{bm}

\newcommand{\figref}[1]{Fig.~\ref{#1}}

\newcommand{\punc}[1]{\,#1}
\newcommand{\neweqnline}{\nonumber\\}

\begin{document}
\title{Microscopic theory of the magnetoresistance of disordered superconducting films}
\author{G.J.~Conduit}
\affiliation{Department of Physics, Ben Gurion University, Beer Sheva 84105, Israel}
\author{Y.~Meir}
\affiliation{Department of Physics, Ben Gurion University, Beer Sheva 84105, Israel}

\date{\today}

\begin{abstract}
  Experiments on disordered superconducting thin films have revealed a
  magnetoresistance peak of several orders of magnitude. Starting from the
  disordered negative-$U$ Hubbard model, we employ an \emph{ab initio}
  approach that includes thermal fluctuations to calculate the resistance,
  and fully reproduces the experimental phenomenology. Maps of the
  microscopic current flow and local potential allow us to pinpoint the
  source of the magnetoresistance peak -- the conducting weak links change
  from normal on the low-field side of the peak to superconducting on the
  high-field side. Finally, we formulate a simple one-dimensional model to
  demonstrate how small superconducting regions will act as weak links in
  such a disordered thin film.
\end{abstract}


\maketitle

The interplay of disorder and superconductivity has been a subject of great
interest and debate for many years \cite{Anderson59,Gorkov59,BCS50}. While
the consequences of disorder are relatively well understood within BCS
mean-field theory, the way disorder modifies the magnetic field and the
temperature response of low-dimensional superconductors (SCRs), where the
loss of superconductivity is due to phase fluctuations \cite{BKT}, remains
an open question. This situation is underscored by recent puzzling
experiments that demonstrated a huge magnetoresistance (MR) peak on the
normal side of the SCR-insulator transition in thin films \cite{MR}, the
emergence of a ``super-insulating'' phase \cite{SI} in the same system, and
the smearing of the transition at the two-dimensional SC interface formed
between two insulating oxides \cite{LaSr}. One of the main reasons for our
limited understanding of these experiments is that there is no microscopic
theory that incorporates disorder, phase fluctuations and magnetic field. In
this Letter we utilize a recently developed \emph{ab initio}
formalism~\cite{ConduitMeir10} to address the puzzle of the MR peak. This
tool allows us to calculate the conductance through a possibly disordered
superconducting (SC) region, based on a microscopic model, using an
approximation that takes into account the phase and amplitude fluctuations
of the SC order parameter. Moreover, the detailed information provided by
this approach on the local currents and chemical potentials (for details
see~\cite{ConduitMeir10}), in addition to the nature of the current flow
(electrons or Cooper pairs) illuminates the microscopic physics behind the
anomalous resistance peak.

The SC region is described by the negative-$U$ Hubbard model, a lattice
model that includes on-site attraction, and may include disorder, orbital
and Zeeman magnetic fields, and Coulomb repulsive interactions. The
Hamiltonian is
\begin{align}
 \hat{H}&=\sum_{i,\sigma}\epsilon_{i}c^{\dagger}_{i\sigma}c_{i\sigma}
 -\!\!\!\sum_{\langle i,j\rangle,\sigma}\left(t_{ij}c^{\dagger}_{i\sigma}c_{j\sigma}
 +t_{ij}^{*}c^{\dagger}_{j\sigma}c_{i\sigma}\right)\neweqnline
 &-\sum_{i}U_{i}c^{\dagger}_{i\uparrow}c^{\dagger}_{i\downarrow}c_{i\downarrow}c_{i\uparrow}
 +\sum_{i\neq j}U^{\text{Coul}}_{ij}\hat{n}_i\hat{n}_j
 \punc{,}
 \label{eq:Hubbard}
\end{align}
where $\epsilon_{i}$ is the on-site energy, $t_{ij}$ the hopping element
between adjacent sites $i$ and $j$, and $U_{i}$ is the onsite two-particle
attraction. Both $|t_{ij}|=t$ and $U_{i}=U$ are taken to be uniform. An
orbital magnetic field can be incorporated into the phases of the hopping
elements $t_{ij}$ (a Zeeman field, manifested as spin-dependent on-site
energies $\epsilon_{i\sigma}$, will not be included here). To account for
disorder, $\epsilon_{i}$ will be drawn from a Gaussian distribution of width
$W$.  The local density is $\hat{n}_i\equiv \sum_\sigma
c^{\dagger}_{i\sigma}c_{i\sigma}$, and the long-range repulsion is given by
the screened Coulomb interaction,
$U^{\text{Coul}}_{ij}=U_{\text{C}}\exp[-r_{ij}/\lambda]a/r_{ij}$, with
$r_{ij}$ the distance between sites $i$ and $j$, and $\lambda=2a$ the
screening length, both measured in units of the lattice constant $a$ (for
computational efficiency we cut off this interaction after four
sites). Unlike, for example, the disordered \emph{XY} model, the
negative-$U$ Hubbard model can lead to a BCS transition, a BKT transition
\cite{BKT}, or to a percolation transition \cite{Erez2010}, and thus this
choice will not limit \emph{a priori} the underlying physical processes. In
order to compare the results of the model to experiments, one needs to
compare the relevant length scales in particular the SC coherence length and
the mean-free path (or localization length). In the calculations presented
below we keep the former (which is governed practically by $U/t$) fixed, and
tune the latter by changing the disorder $W/t$.

In Ref.\cite{ConduitMeir10} we have developed an exact formula for the
current through such a SC region, sandwiched between metallic
leads. Furthermore, the resulting expression for the current can be
evaluated using a Monte Carlo approach that treats the fermionic degrees of
freedom exactly, takes into account the thermal fluctuations of the
amplitude and phase of the SC order parameter, but neglects its quantum
fluctuations. It has been demonstrated \cite{ConduitMeir10} that such an
approach describes quantitatively various transport processes in the
vicinity of the BKT transition, which are crucial to describe the physics of
low-dimensional superconductors.

\begin{figure}
 \centerline{\resizebox{0.95\linewidth}{!}{\includegraphics{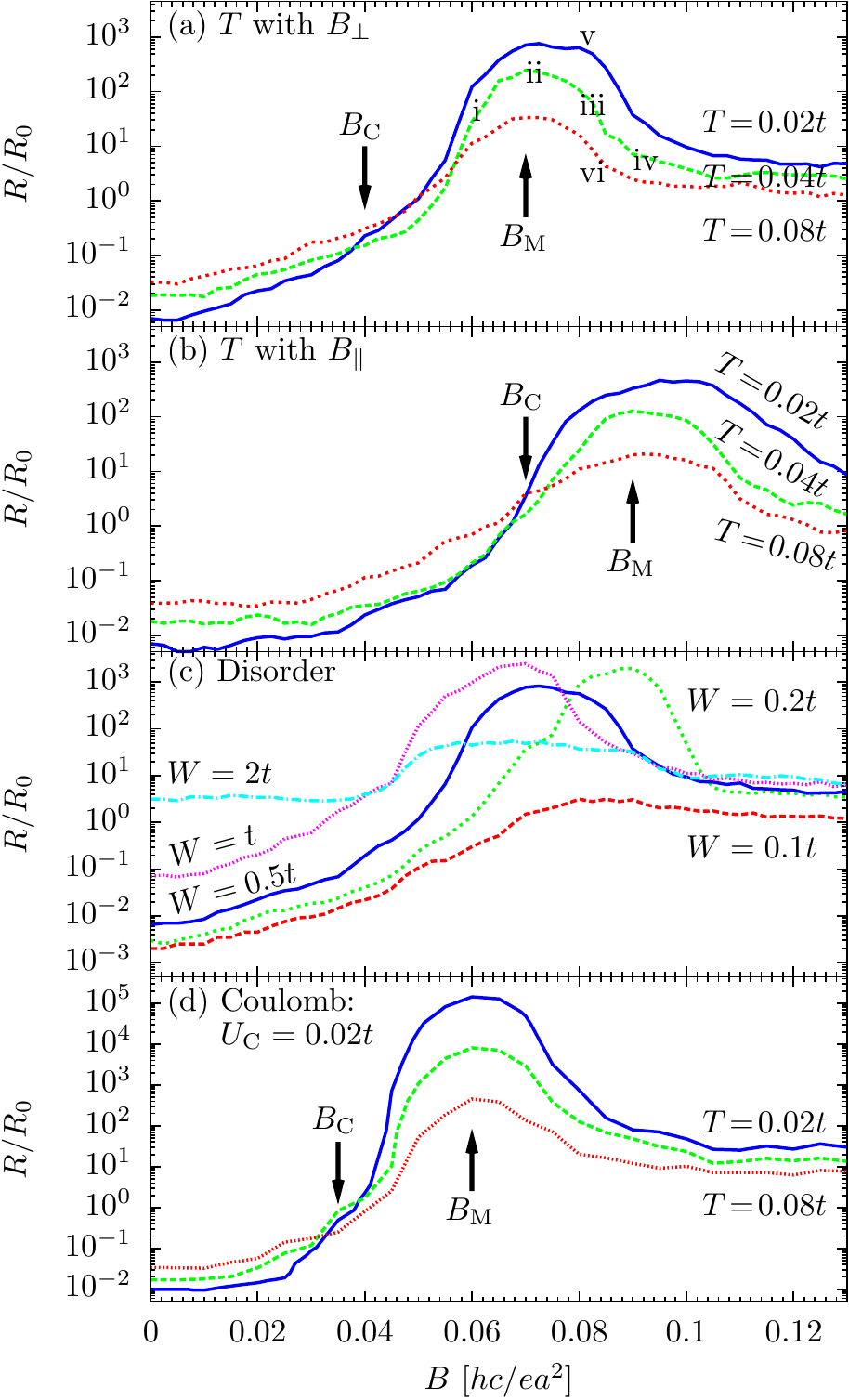}}}
 \caption{(Color online) (a,b) The MR with an applied magnetic field (a)
   perpendicular to the surface and (b) parallel to the surface. The blue
   curve (highest at large fields) is at low temperature, yellow
   intermediate and red high temperature. The arrows highlight the critical
   field $B_{\text{C}}$ and the maximum resistance field $B_{\text{M}}$.
   The points (i - vi) denote the magnetic fields where the current maps of
   \figref{fig:CurrentMaps} were evaluated.  The perpendicular MR for four
   different amplitudes of disorder is shown in (c), and with Coulomb
   repulsion in (d).  }
 \label{fig:DisorderRealizationsVary}
\end{figure}

In the following we first recover the main experimental results and study
how the microscopic variables affect the resistance.  We then construct a
detailed map of the current flow through the system to uncover the
microscopic mechanism that drives the MR peak. First, in
\figref{fig:DisorderRealizationsVary}(a) we depict the resistance at several
temperatures for a sample of size $L_{\text{x}}=21a$ ($\hat{x}$ is the
direction of the applied potential), $L_{\text{y}}=11a$, and
$L_{\text{z}}=2a$. The interaction strength is $U=2t$, the onsite disorder
has width $W=0.5t$, the average density is at $41\%$ filling, and initially
there are no long-range repulsive interactions, $U_{\text{C}}=0$. Throughout
we set $k_{\text{B}}=1$. At zero magnetic field, the resistance of the SC
state is solely the lead-SCR contact resistance. Increasing the magnetic
field perpendicular to the surface drives a SCR-insulator transition, around
$B_{\text{C}}\simeq0.04$, in units of (electron) flux quantum per square
$[hc/ea^{2}]$, manifested in the reversal of the temperature dependence of
the resistance. Above the transition the system displays, in agreement with
experiment \cite{MR}, a peak resistance (at the field
$B_{\text{M}}\simeq0.07$) that is over a hundred times the normal state
resistance. This feature consistently emerges for several different
statistical configurations of disorder, with the peak resistance varying by
a factor of approximately ten. On changing the direction of the (orbital)
magnetic field from normal to lying in the plane of the sample
(\figref{fig:DisorderRealizationsVary}(b)), both $B_{\text{C}}$ and
$B_{\text{M}}$ shift to higher fields, with a slight decrease in the
magnitude of the MR peak, in complete agreement with recent experiments
\cite{Johansson11}.

\begin{figure}
 \centerline{\resizebox{1.0\linewidth}{!}{\includegraphics{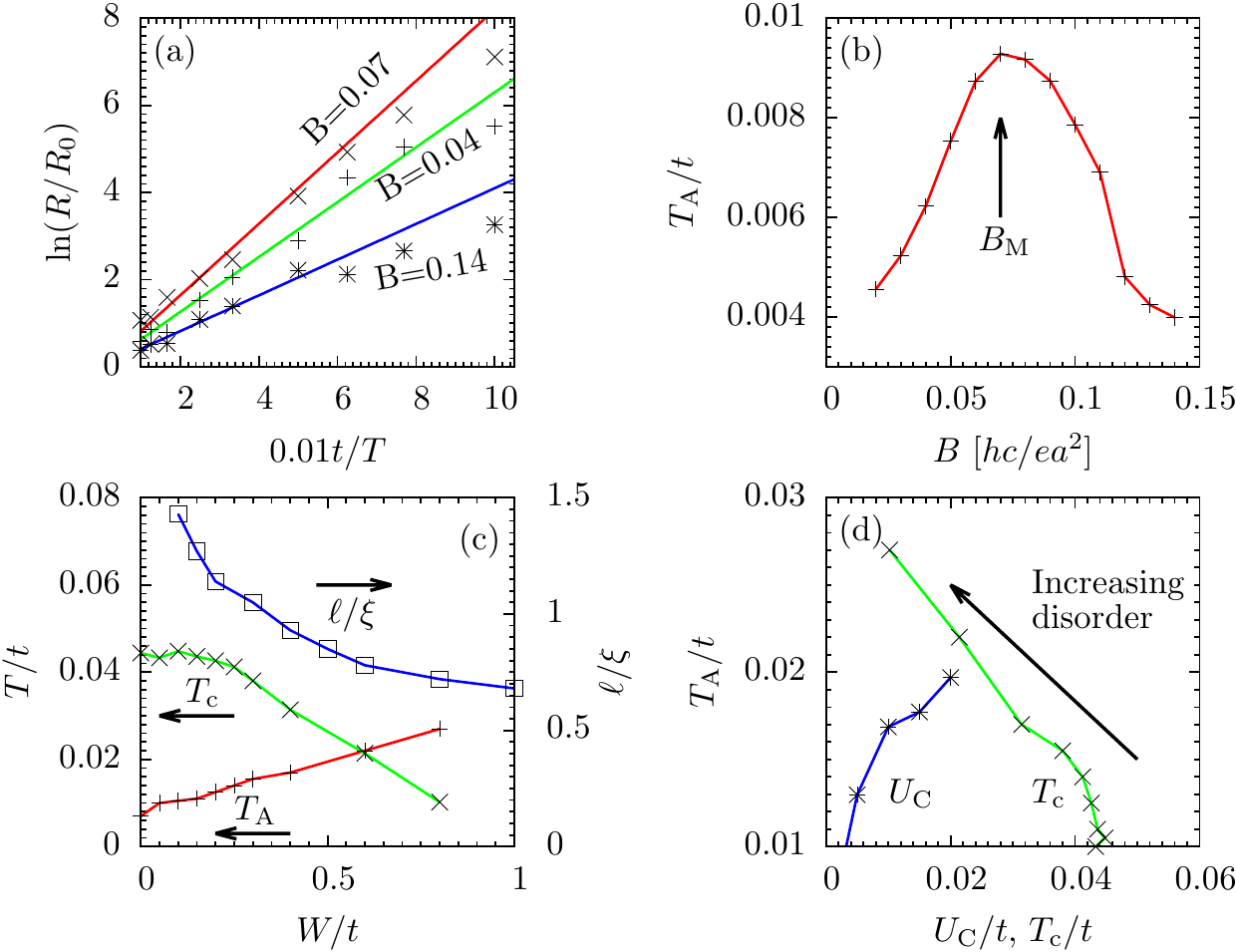}}}
 \caption{(Color online)
 (a) The temperature dependence of the resistance at three values of
 the magnetic field.
 (b) The activation energy $T_{\text{A}}$ with magnetic field.
 (c) The activation energy $T_{\text{A}}$
 (pluses, red), superconducting transition temperature $T_{\text{c}}$
 (crosses, green), and ratio of mean free path $\ell$ to coherence
 length $\xi$ (squares, blue) with disorder.
 (d) $T_{\text{A}}$ vs $T_{\text{c}}$ and
 screened Coulomb interaction $U_{\text{C}}$ for different disorder levels.
 }\label{fig:ActivationTemp}
\end{figure}

We now explore how the MR varies with the parameters of the system.
Firstly, in \figref{fig:DisorderRealizationsVary}(c) we focus on one
specific realization of disorder but increase its amplitude. At low
disorder, $W=0.1t$, the magnetic field suppresses the SC state, but there is
no MR peak. On increasing the magnitude of the disorder to $W=0.2t$ a peak
in the resistive curve emerges at $B\approx0.09$. (Interestingly, the peak
emerges approximately when the mean free path $\ell$ becomes smaller than
the coherence length $\xi$, \figref{fig:ActivationTemp}(c).) As disorder is
further increased both the SC transition and MR peak shift to lower magnetic
fields~\cite{Siemons08}. In agreement with experiment, the MR peak persists even when the SC
phase is suppressed at zero field ($W=0.5t$ and $W=t$), until the peak is
extinguished at large enough disorder
($W=2t$). \figref{fig:DisorderRealizationsVary}(d) demonstrates that the MR
peak becomes more dramatic with increasing Coulomb repulsion, as the MR peak
increases by two orders of magnitude. The enhancement of the MR peak by both
disorder and Coulomb repulsion points towards a Coulomb blockade mechanism.

\begin{figure*}
 \begin{tabular}{@{}l@{\quad}l@{\quad}l@{}}
  (i) $B=0.06hc/ea^{2}$, $T=0.04t$&(ii) $B=0.07hc/ea^{2}$, $T=0.04t$&(iii) $B=0.08hc/ea^{2}$, $T=0.04t$\\
  \resizebox{0.32\linewidth}{!}{\includegraphics{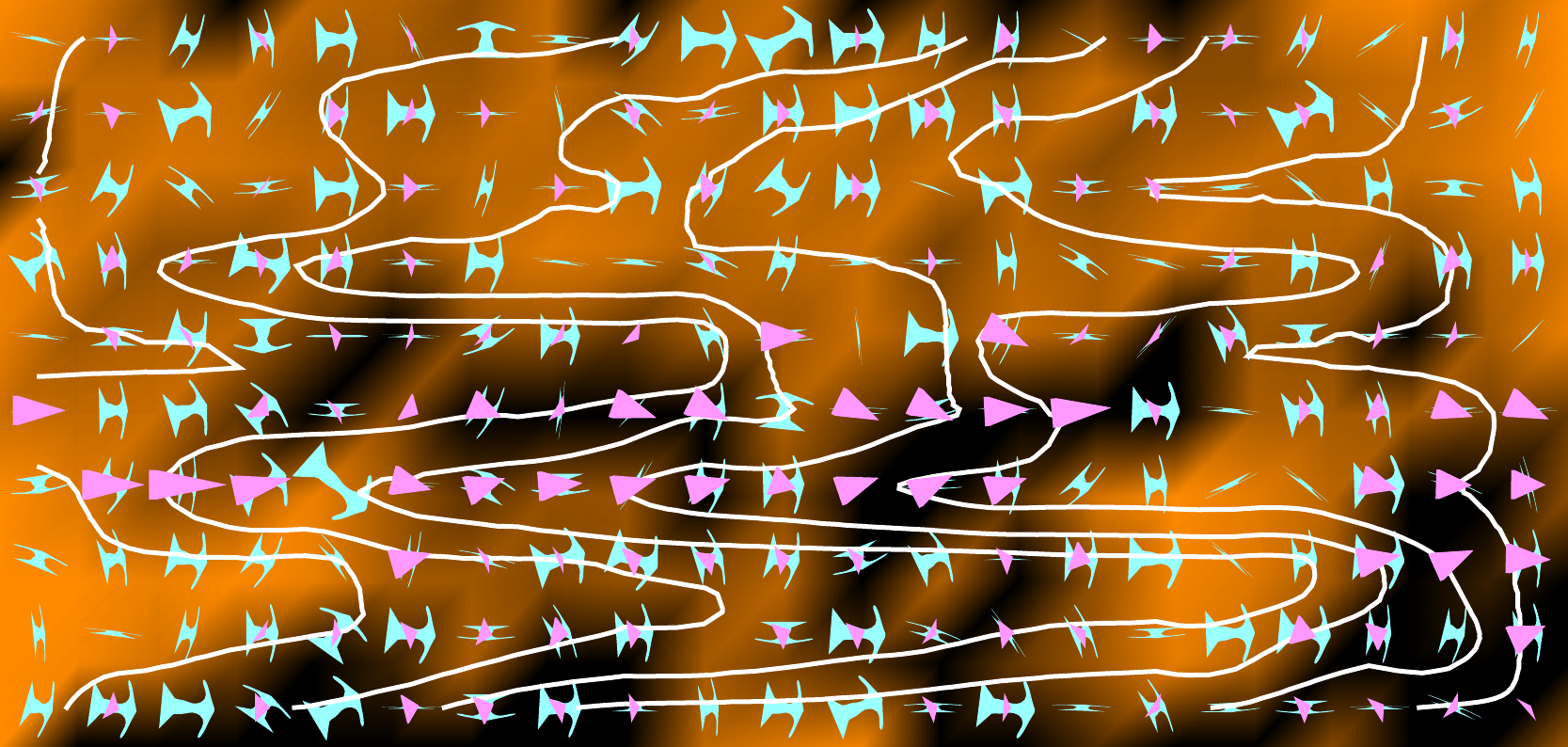}}&
  \resizebox{0.32\linewidth}{!}{\includegraphics{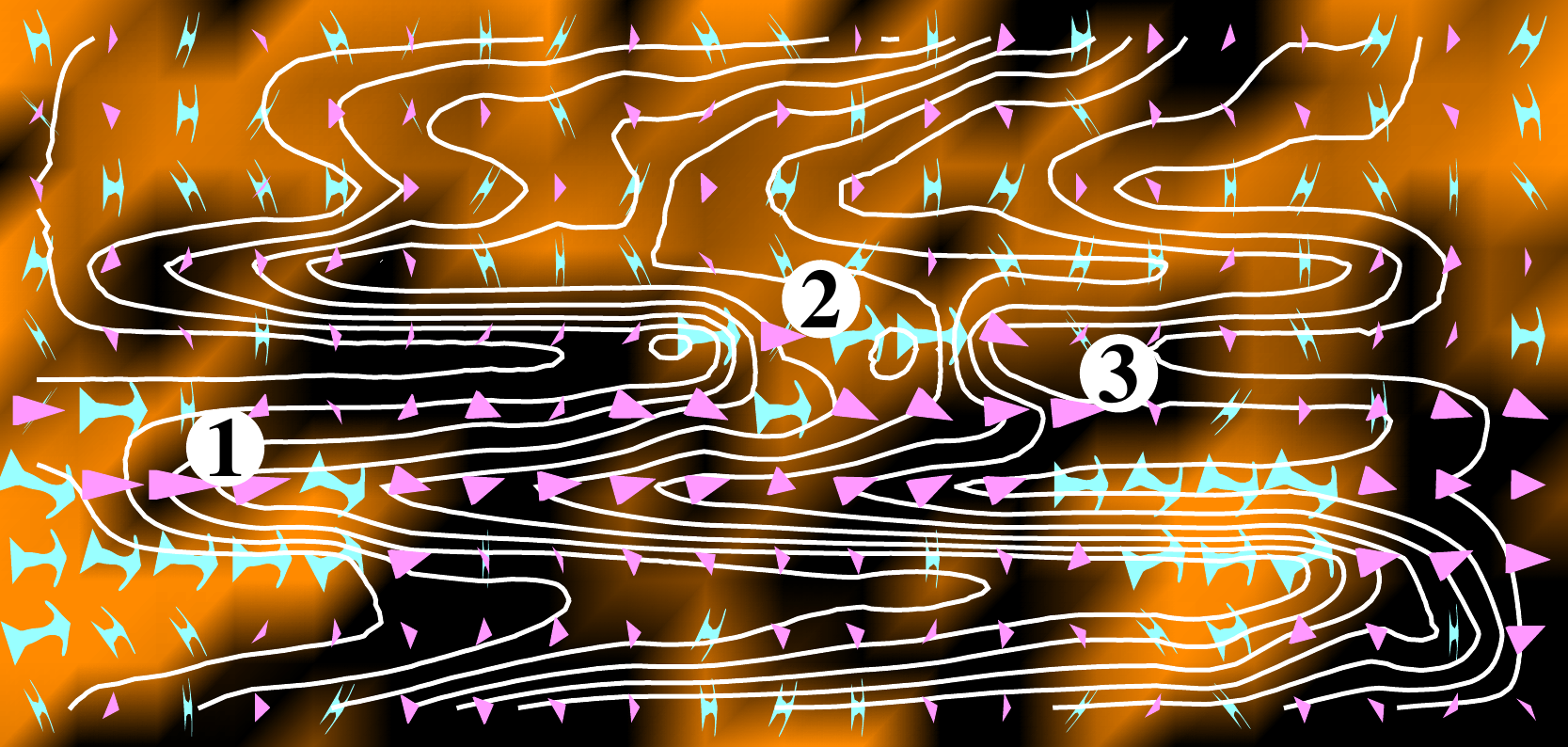}}&
  \resizebox{0.32\linewidth}{!}{\includegraphics{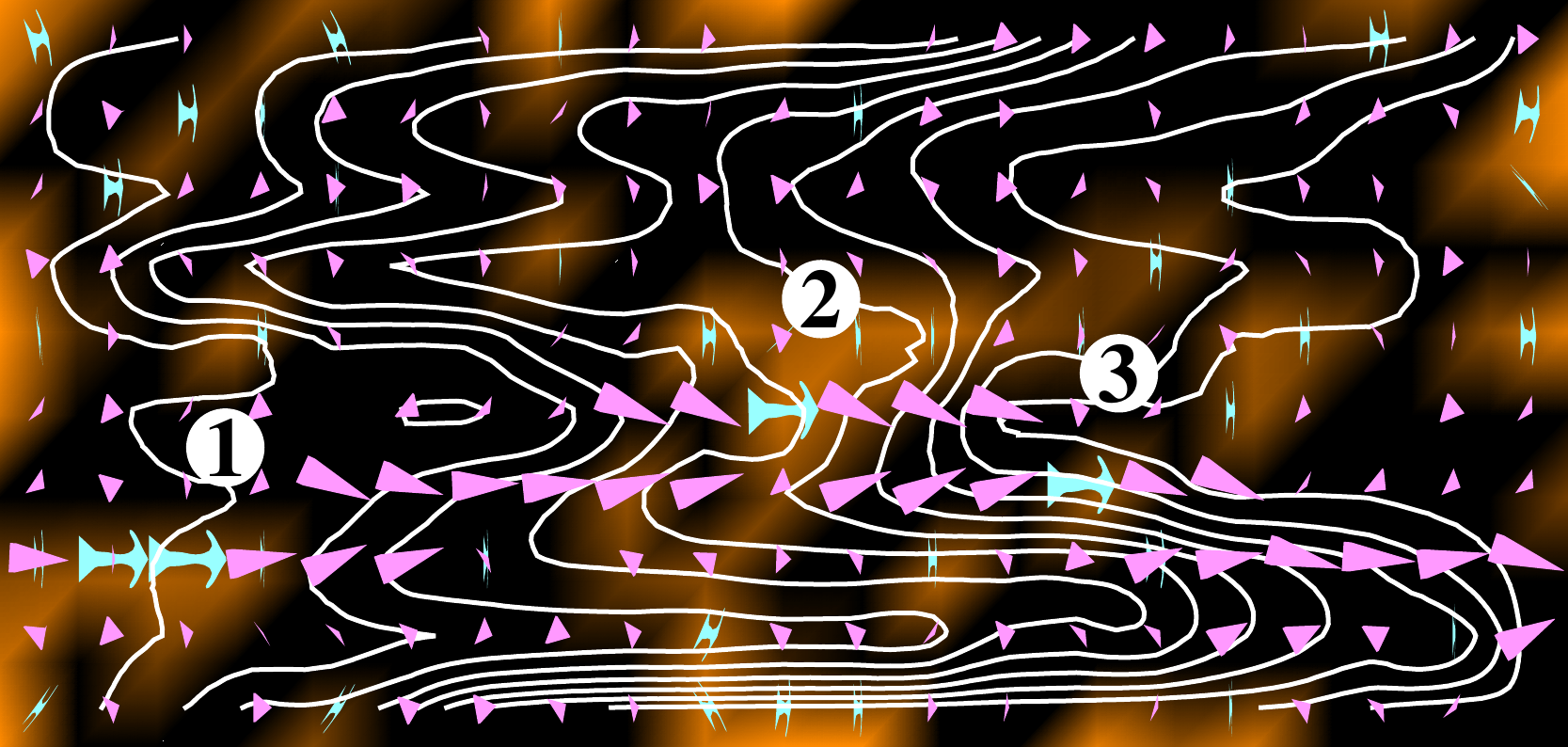}}\\
  (iv) $B=0.09hc/ea^{2}$, $T=0.04t$&(v) $B=0.08hc/ea^{2}$, $T=0.02t$&(vi) $B=0.08hc/ea^{2}$, $T=0.08t$\\
  \resizebox{0.32\linewidth}{!}{\includegraphics{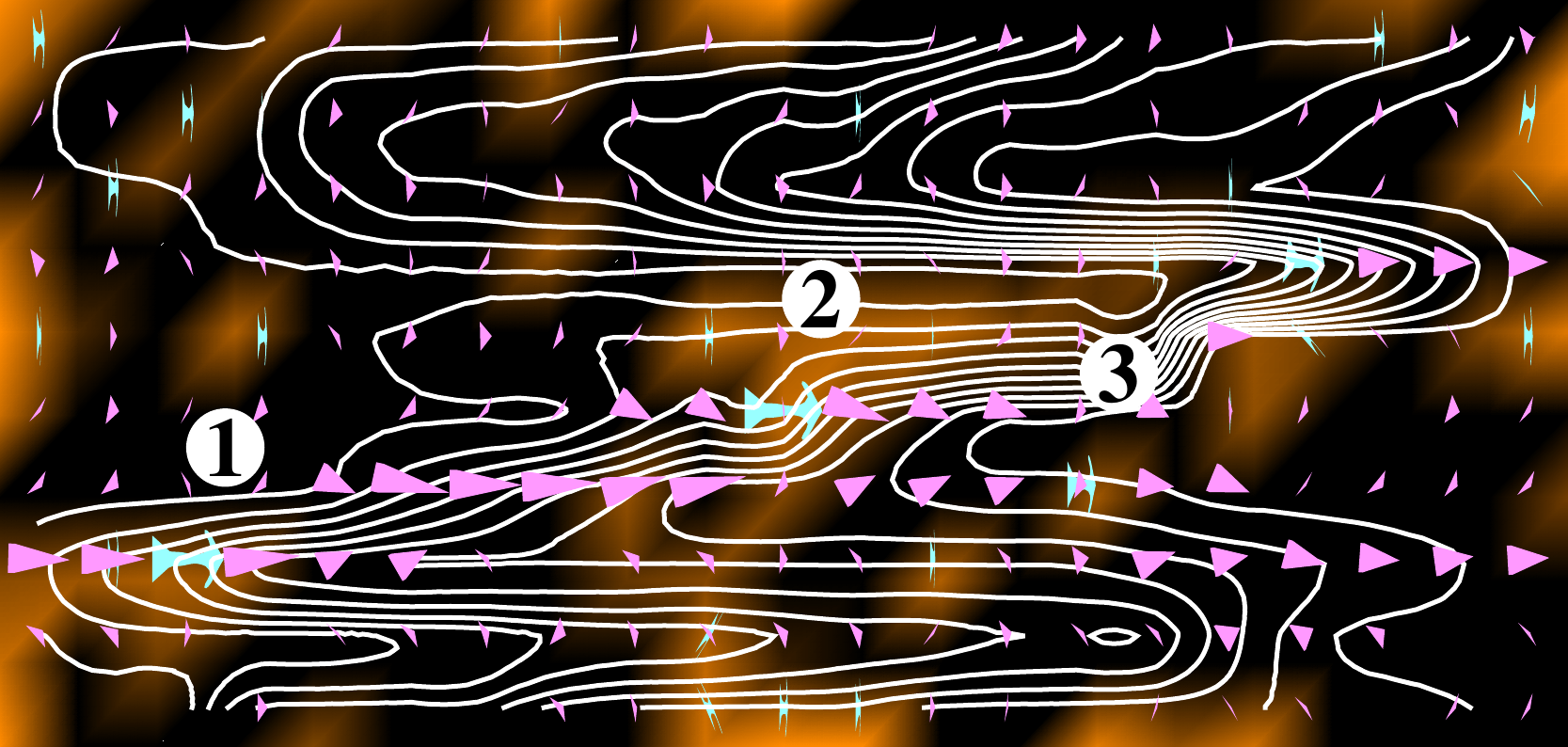}}&
  \resizebox{0.32\linewidth}{!}{\includegraphics{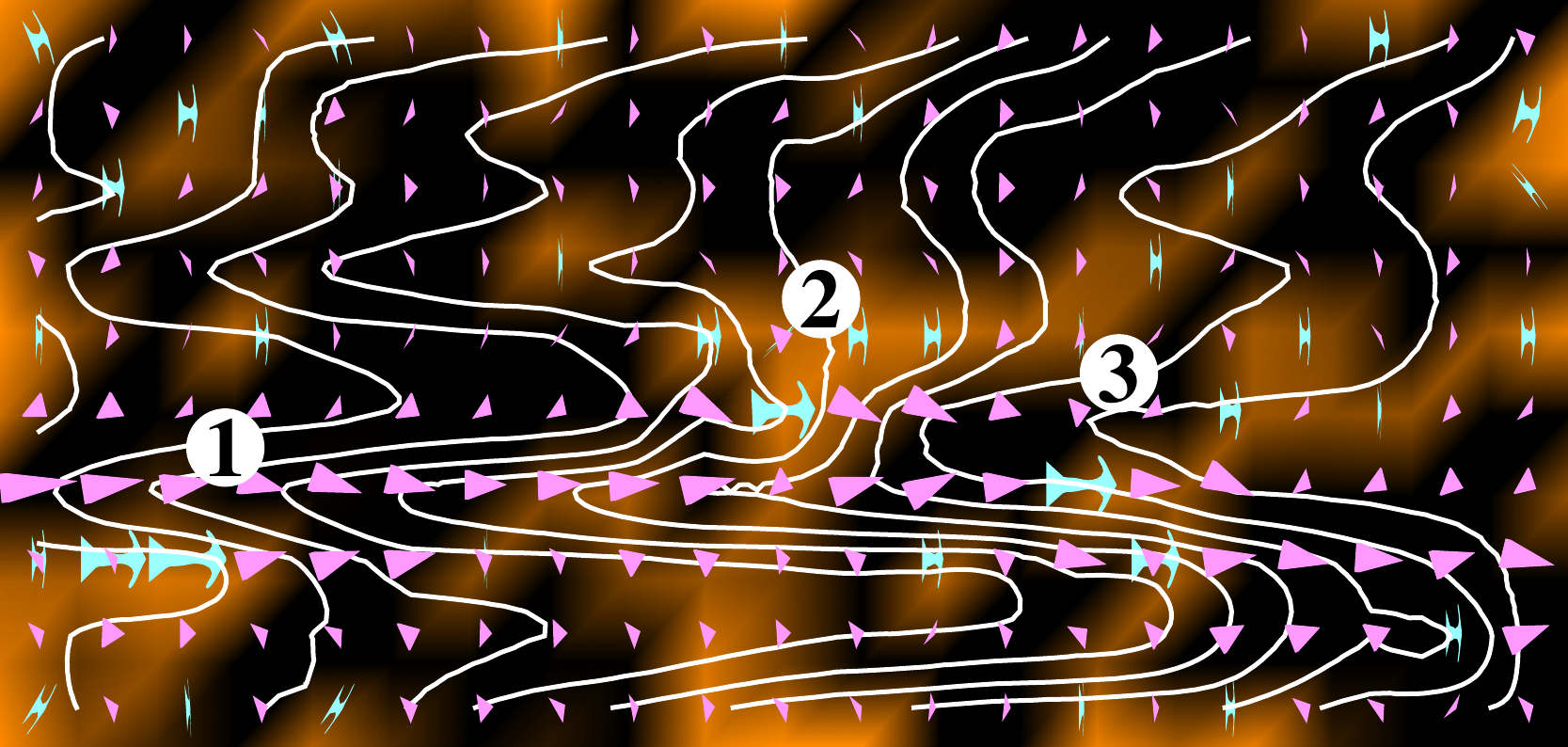}}&
  \resizebox{0.32\linewidth}{!}{\includegraphics{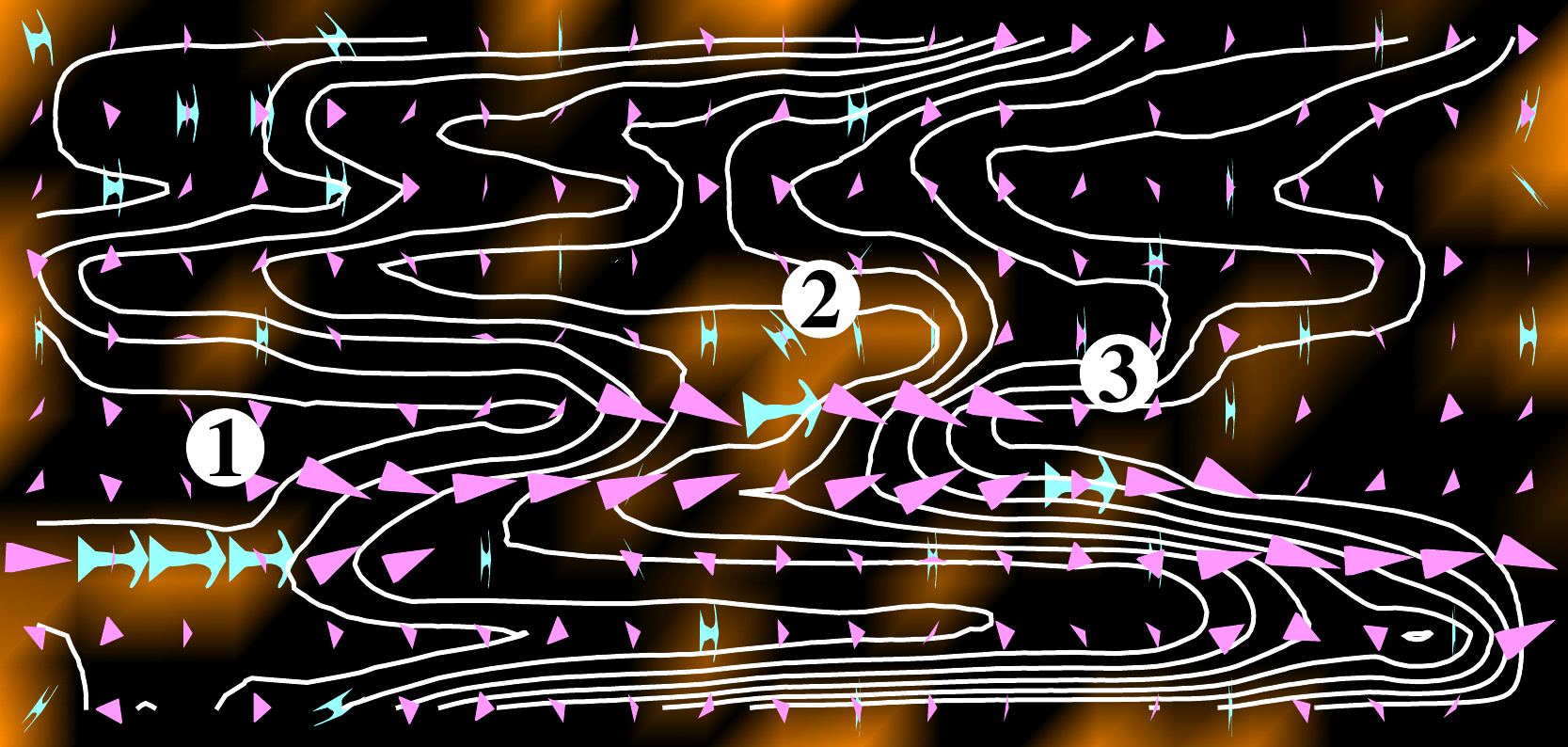}}
 \end{tabular}
 \caption{(Color online) Current maps at six values of the normal magnetic
   field shown in \figref{fig:DisorderRealizationsVary}(a). The net current
   flow is from left to right. The cyan darts show supercurrent and magenta
   pointers normal current direction and magnitude; their magnitude is
   renormalized in each map according to the maximum local current flow. The
   white lines show equi-chemical potential lines across the sample (ten
   lines correspond to the full potential drop between the two leads), and
   the background shading depicts the magnitude of the local superconducting
   order parameter ranging from strong (orange) to zero (black). The labels
   (1) to (3) highlight three points of special interest.}
 \label{fig:CurrentMaps}
\end{figure*}

\figref{fig:ActivationTemp}(a) depicts the temperature dependence of the
resistance for several values of magnetic field, which demonstrate an
activated behavior, $R=R_{0}\exp(T_{\text{A}}/T)$. The activated behavior,
and the peak of the activation temperature $T_{\text{A}}$ at the field
$B_{\text{M}}$ (\figref{fig:ActivationTemp}(b)) agrees with experimental
observations~\cite{MR}. $T_{\text{A}}$ increases with disorder
(\figref{fig:ActivationTemp}(c)) and with repulsive interactions
(\figref{fig:ActivationTemp}(d)), consistent with the enhancement of the MR
peak by these parameters (\figref{fig:DisorderRealizationsVary}(c) and (d)).
The critical temperature $T_{\text{C}}$ at which SC correlations are
suppressed at zero-field on the other hand falls with disorder. In
\figref{fig:ActivationTemp}(d) this gives us an inverse linear dependence of
$T_{\text{A}}$ on $T_{\text{C}}$, similar to that seen in
experiment~\cite{MR}.

So far we have demonstrated that the calculation reproduces the
phenomenology of the experimental observations and hinted that activated
transport through a blockade region drives the rise of the resistance. We
now take advantage of our ability to probe the evolution of the local
currents and chemical potentials with applied magnetic field
(\figref{fig:CurrentMaps} and comprehensively in Fig.~S2) to pinpoint the
physical processes behind this behavior. Below the critical field,
$B_{\text{C}}=0.04$, Fig.~S2 shows that no voltage drops across the sample,
though the current flow, consisting of Cooper pairs, becomes nonuniform. As
the field increases, ($B=0.06$, \figref{fig:CurrentMaps}(i)), there is no
longer long-range SC coherence across the system, and current crossing the
system changes its nature between SC and normal. Above this field the
current has to traverse normal areas of the sample, and as their density
increases with magnetic field, the resistance rises. As the field increases
further ($B=0.07$, Panel (ii)), the disorder induces specific channels of
transport -- the current starts as SC, changes into normal current around
point (1) and then splits towards points (2) and (3) where it reverts to
Cooper pairs, only to change back into normal current as it enters the
right-hand lead. At this magnetic field, near the maximal value of the
resistance, there are approximately equal contributions to the current from
electrons and from Cooper pairs. With increasing field ($B=0.08$, Panel
(iii)) the SC areas (2) and (3) shrink and become the main source of voltage
drop and resistance in the sample. A larger field ($B=0.09$, Panel (iv))
suppresses the SC correlations, lowering the resistance of the weak links
and the overall the resistance. This decrease is, in fact, an interplay of
two phenomena -- larger SC areas do not serve as weak links (see point (1)
in \figref{fig:CurrentMaps}(iii)), but as they shrink with increasing
magnetic field the resistance associated with them increases significantly
(see point (1) in \figref{fig:CurrentMaps}(iv)). At the same time small SC
areas, that gave a significant contribution to the resistance (point (3) in
these two panels) become normal and the overall resistance decreases. A much
larger magnetic field (Fig.~S2) suppresses superconductivity almost
completely, giving rise to a uniform drop of voltage across the sample and a
significantly lower resistance. Similar behavior has been obtained in the
presence of repulsive Coulomb interactions (not shown).

To explore the origin of the Arrhenius behavior panels (v) and (vi) depict
the current flow at $B=0.08$ at lower and higher temperatures than Panel
(iii). At point (1) we see direct evidence of the activated transport: on
lowering the temperature (Panel (v)) more potential is dropped across the
weak SC link, whereas with increasing temperature (Panel (vi)) the potential
dropped across this weak link falls. The effect of the weak link is so
profound that in the hotter panel (vi) despite the reducing SC order more
supercurrent flows as the electrons no longer skirt around the SC island as
they did in Panel (v).  Interestingly at other places in the sample we see
the more conventional effect of temperature suppressing superconductivity,
for example at point (3) the increasing temperature from Panel (v) to Panel
(iii) reduces the SC order parameter and the supercurrent.

\begin{figure}
 \centerline{\resizebox{1.0\linewidth}{!}{\includegraphics{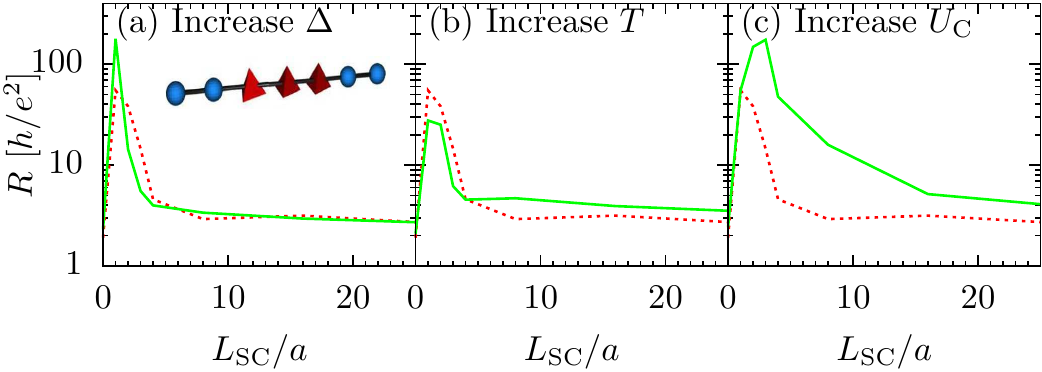}}}
 \caption{(Color online) The inset of (a) shows the 1D model of a normal
   state (blue spheres) conducting channel with weak disorder containing a
   superconducting grain (red prisms) of varying length $L_{\text{SC}}$.  In
   (a) to (c) the red dotted curve is for the same set of standard
   parameters, and the green solid curves correspond to (a) higher dot order
   parameter $\Delta=0.9t$, (b) higher temperature $T=0.1t$, and (c) with
   Coulomb repulsion $U_{\text{C}}=0.02$.  }\label{fig:DotModel}
\end{figure}

Why would small SC areas serve as weak links and contribute to the
increasing resistance? To shed light on this question we studied a
toy-model, consisting of a SC island with finite order parameter
$\Delta=0.3t$ of length $L_{\text{SC}}$, embedded in disordered normal chain
with $\Delta=0$, equal nearest neighbor tunneling, and a temperature
$T=0.05t$, shown in the inset of \figref{fig:DotModel}(a). The main figure
depicts the resistance of the chain as a function of $L_{\text{SC}}$, with
the total chain length fixed. As the SC region becomes shorter, there is a
mild increase in the resistance due the increasing length of the resistive
normal regions. However, when the SC segment is shorter than $4a$ (similar
in size to the SC islands around the resistance peak in
\figref{fig:CurrentMaps}(iii)), there is a sudden increase in the resistance
of the sample. The resistance peak is enhanced by raising the SC order to
$\Delta=0.9t$, (\figref{fig:DotModel}(a)), or increasing temperature to
$T=0.1t$ (\figref{fig:DotModel}(b)), demonstrating an activated transport
process. This is enhanced on approaching the Anderson limit, where the
single particle level-spacing in the SC segment becomes larger than $\Delta$
\cite{Anderson}. Further corroboration of this picture comes from
\figref{fig:DotModel}(a) where the resistance peak emerges at shorter
$L_{\text{SC}}$ for higher $\Delta$. Introducing repulsive interactions
$U_{\text{C}}=0.05t$ (\figref{fig:DotModel}(c)) has two major effects --
there is a larger increase in the resistance and the effect starts at much
larger SC segment size, because now the SC gap has to be compared to the
Coulomb blockade energy rather than the single-particle spacing. Note that
this happens even though this segment is not weakly coupled to the rest of
the chain. This is consistent with the description of the MR peak in the
two-dimensional systems, where as the SC islands shrink the level spacing,
or, more realistically the Coulomb blockade energy becomes of the order of
the SC gap, the islands become the weak links and drive up the resistance of
the sample.

To conclude, we have demonstrated the emergence of a magnetoresistance peak,
starting from a microscopic negative-$U$ Hubbard model. The resistance peak
is driven by the competition between Cooper pair and electron transport,
where on the weak field side of the peak the normal regions serve as the
weak links across the sample, while this role is played by the SC islands on
the high field side of the peak, where Coulomb repulsion probably plays a
dominant role. These ideas can be further tested by checking the effect of
screening by a parallel metallic gate on the peak characteristics. Our
picture has some similarities to the heuristic model presented in
Ref.\cite{Dubi}, though there the resistance on the high-field side of the
peak was completely associated with the normal electrons, while it is
different than the phenomenological approach of
Refs.\cite{Galitski2005,Pokrovsky2010} that discussed the interplay between
quasi-particle and Cooper pair transport.

{\it Acknowledgments:} GJC acknowledges the financial support of the Royal
Commission for the Exhibition of 1851, and the Kreitman Foundation. This
work was also supported by the ISF.

\clearpage

\section{Supplementary material for ``Microscopic theory of the magnetoresistance of
disordered superconducting films''}

\begin{figure}
 \centerline{\resizebox{0.95\linewidth}{!}{\includegraphics{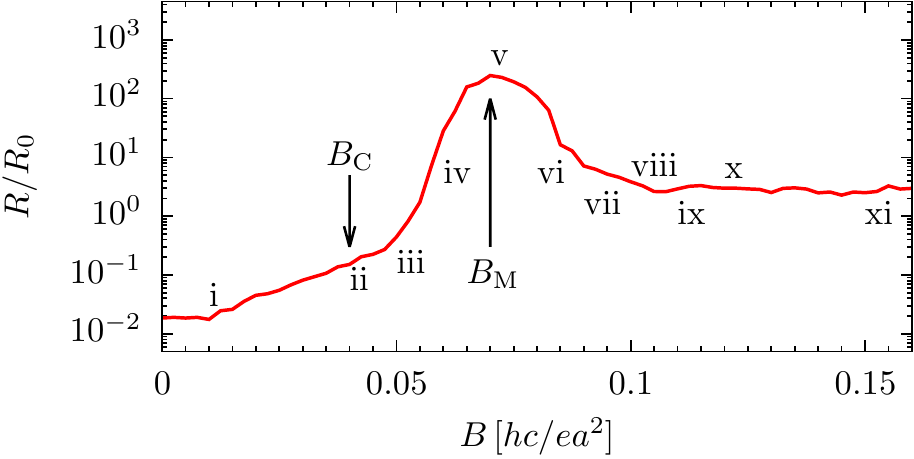}}}
 \caption{The variation of resistance with an
 applied magnetic field normal to the surface at $T=0.04t$. $R_{0}=h/e^{2}$
 is the resistance quantum. The points (i - xi) denote the
 magnetic fields where the current maps of \figref{fig:Maps} were evaluated,
 $B_{\text{C}}$ represents the magnetic field where the temperature
 dependence of the resistance changes sign and $B_{\text{M}}$ the magnetic
 field at the resistance maximum.}
 \label{fig:Resistance}
\end{figure}

\begin{figure*}
 \begin{tabular}{@{}l@{\quad\quad}l@{}}
  (i) $B=0.01hc/ea^{2}$&(ii) $B=0.04hc/ea^{2}$\\
  \resizebox{0.35\linewidth}{!}{\includegraphics{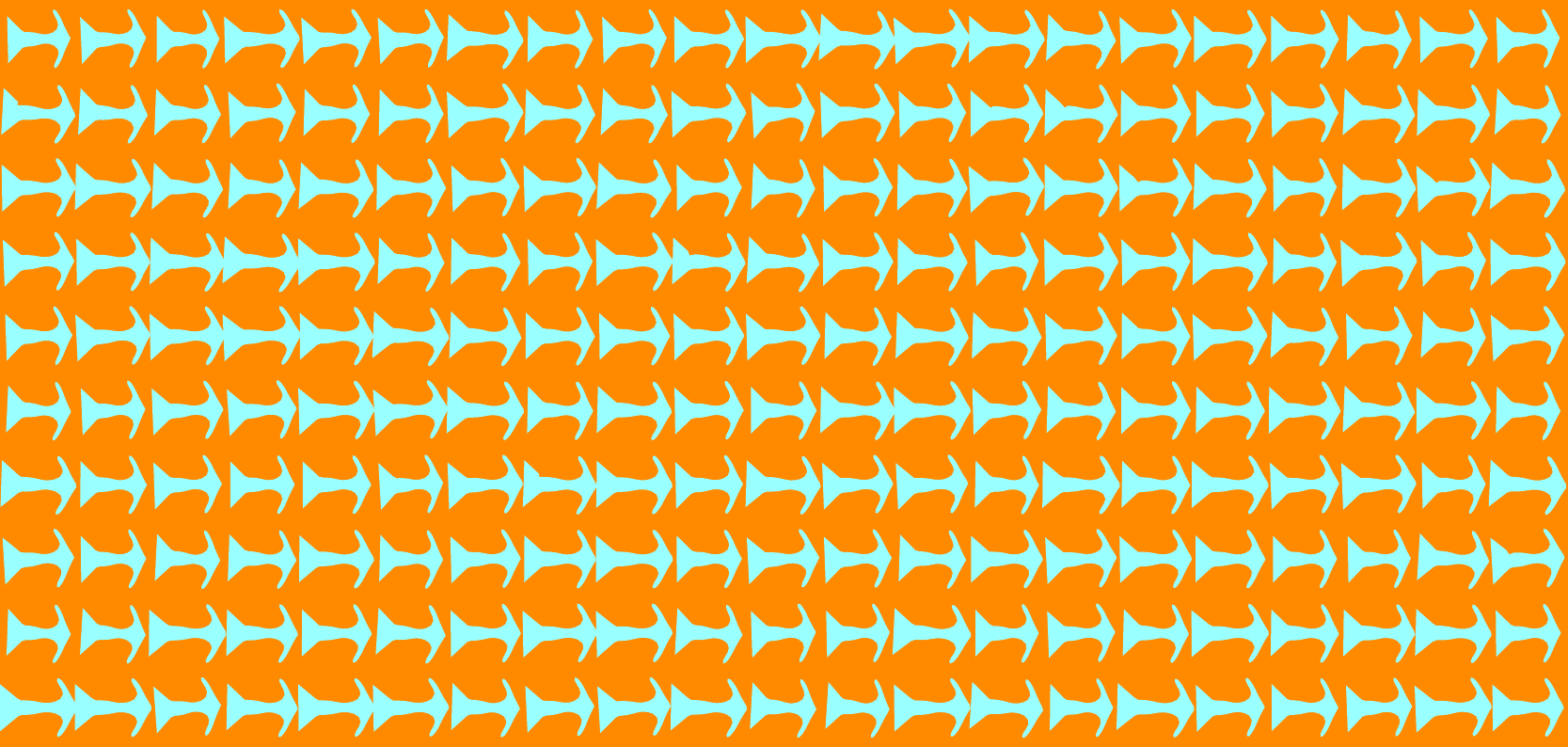}}&
  \resizebox{0.35\linewidth}{!}{\includegraphics{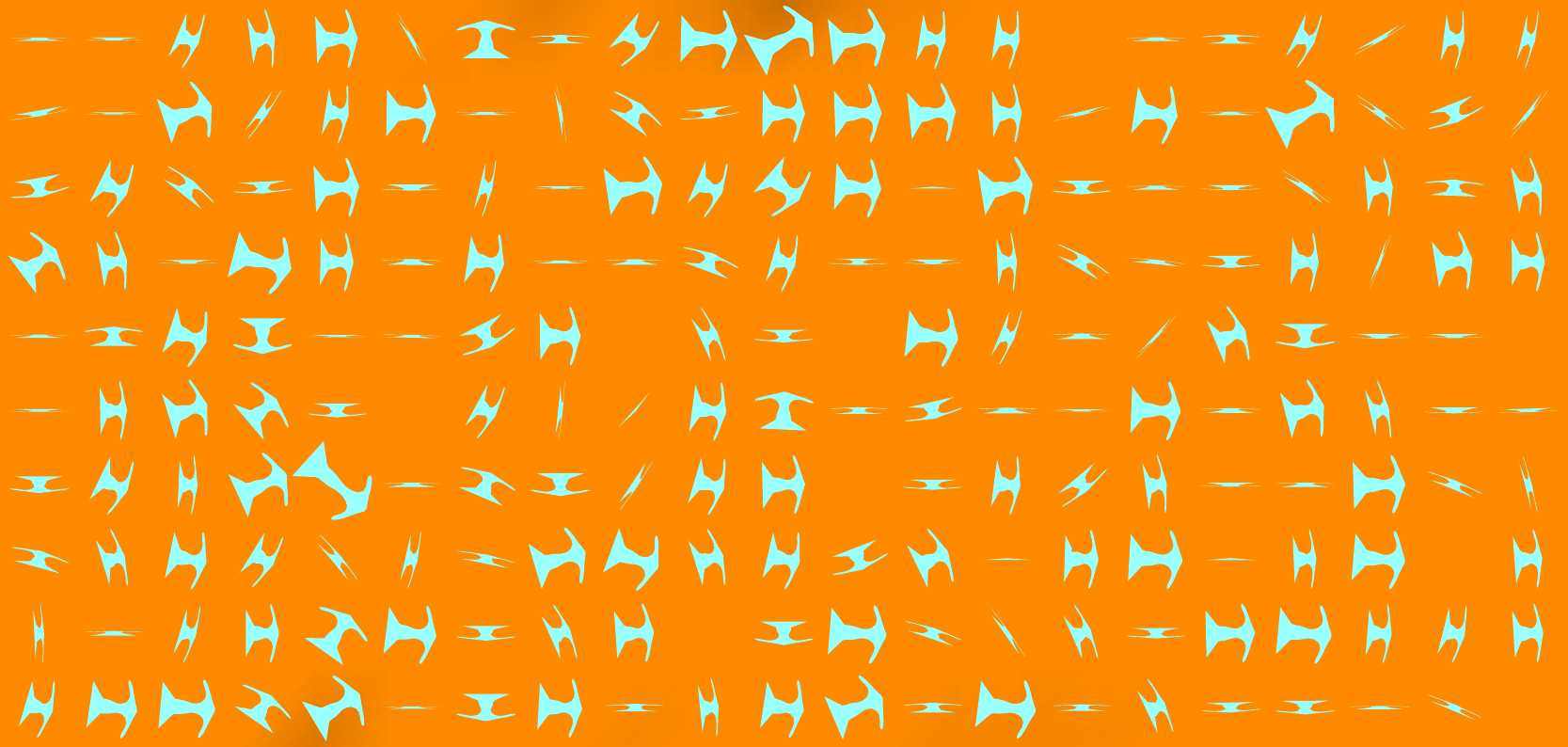}}\\
  (iii) $B=0.05hc/ea^{2}$&(iv) $B=0.06hc/ea^{2}$\\
  \resizebox{0.35\linewidth}{!}{\includegraphics{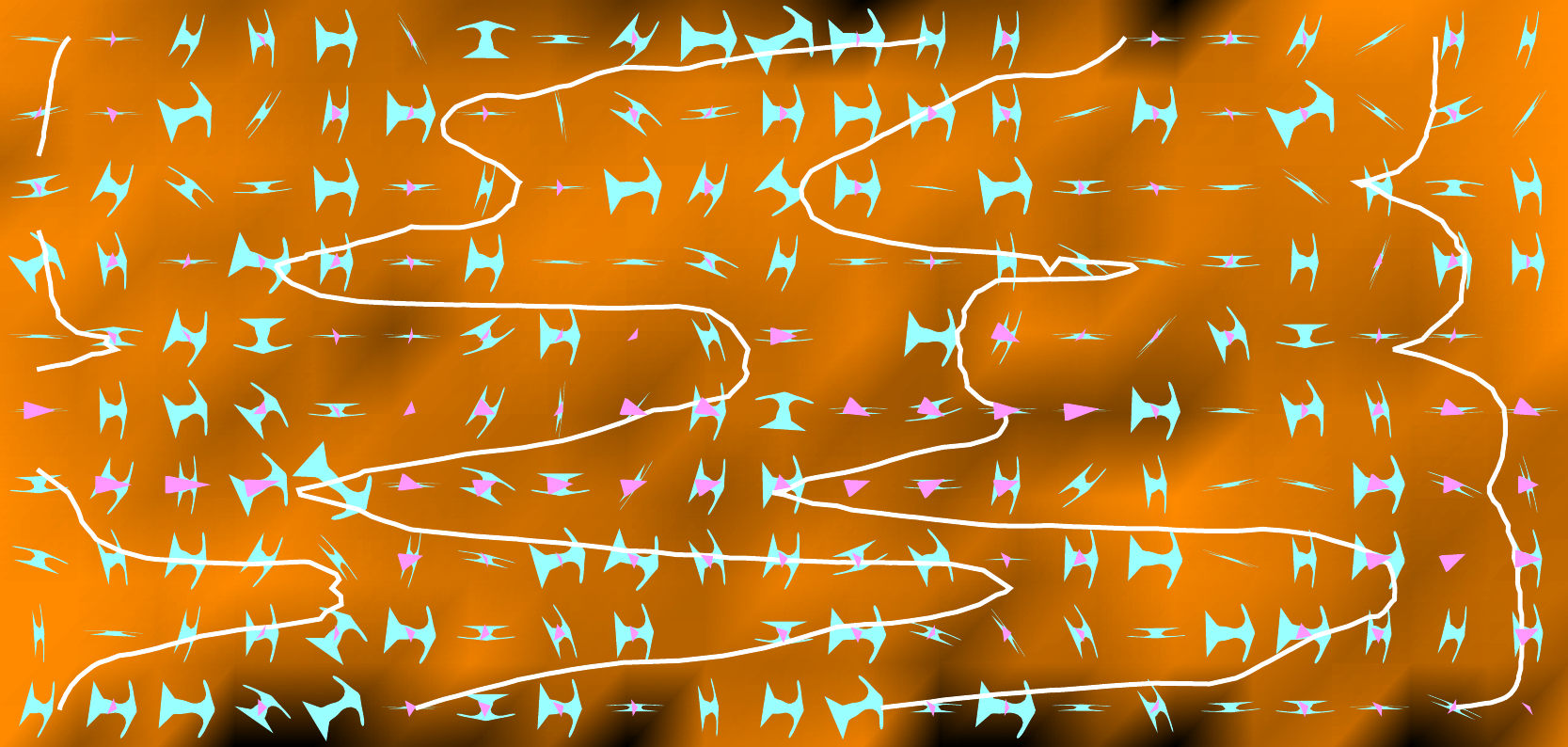}}&
  \resizebox{0.35\linewidth}{!}{\includegraphics{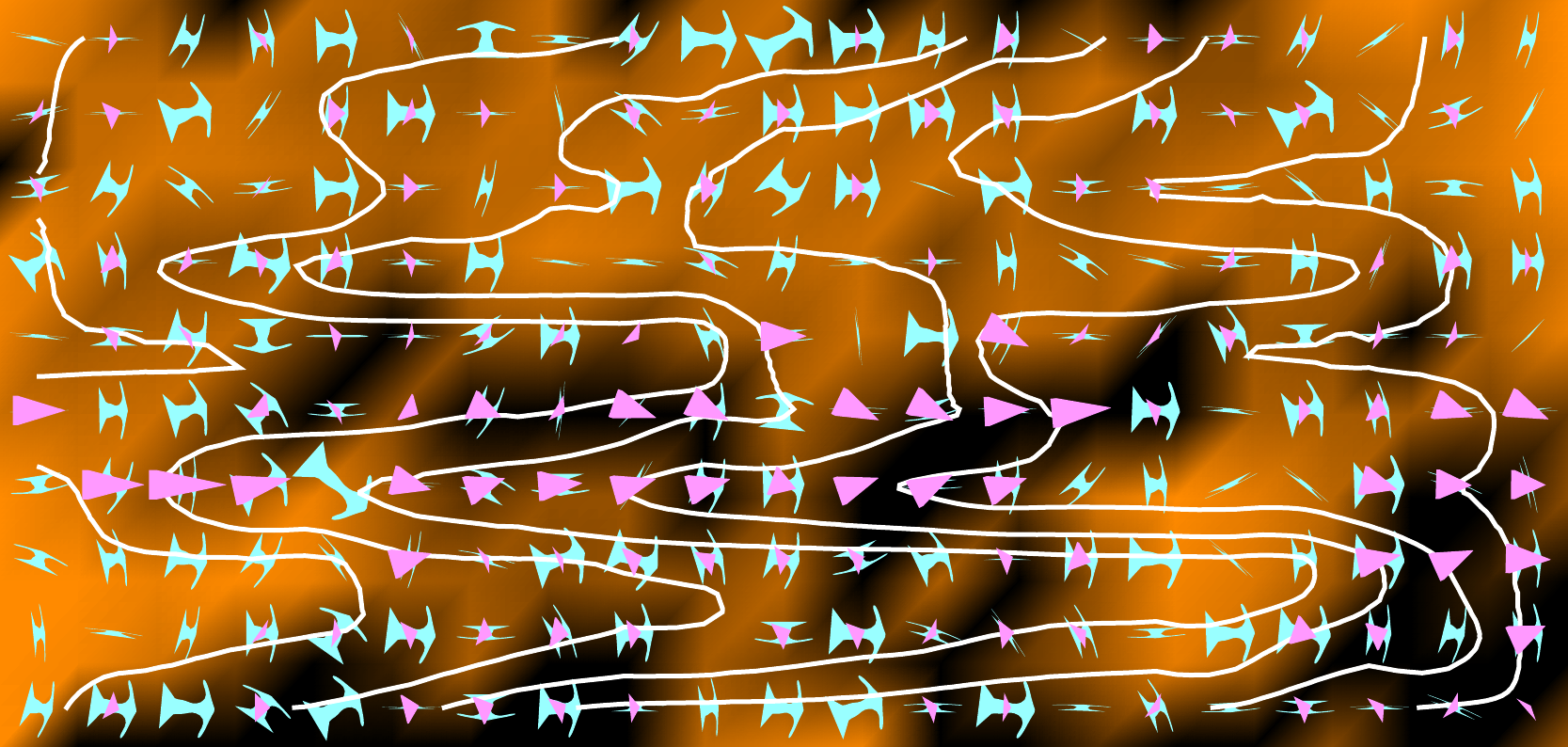}}\\
  (v) $B=0.07hc/ea^{2}$&(vi) $B=0.08hc/ea^{2}$\\
  \resizebox{0.35\linewidth}{!}{\includegraphics{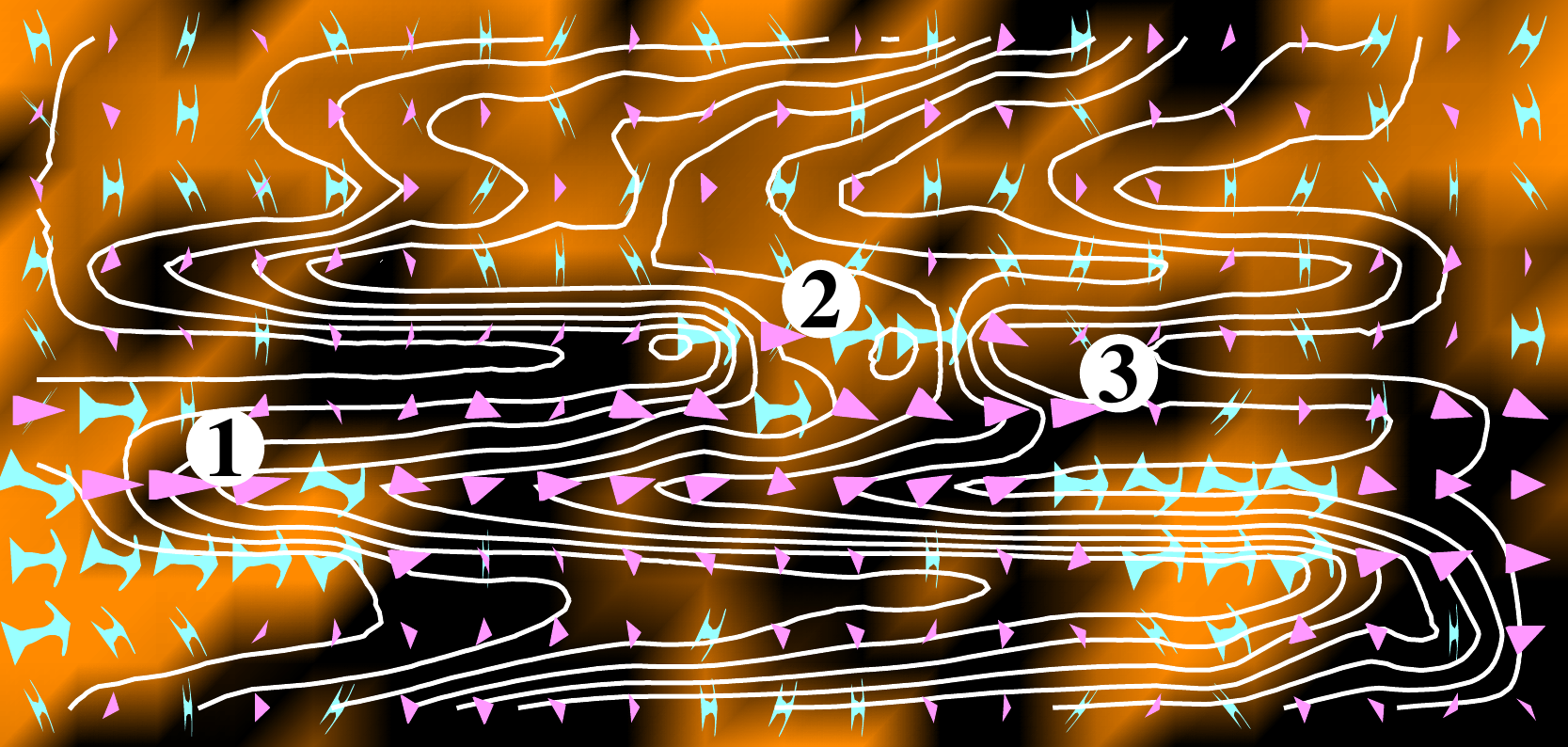}}&
  \resizebox{0.35\linewidth}{!}{\includegraphics{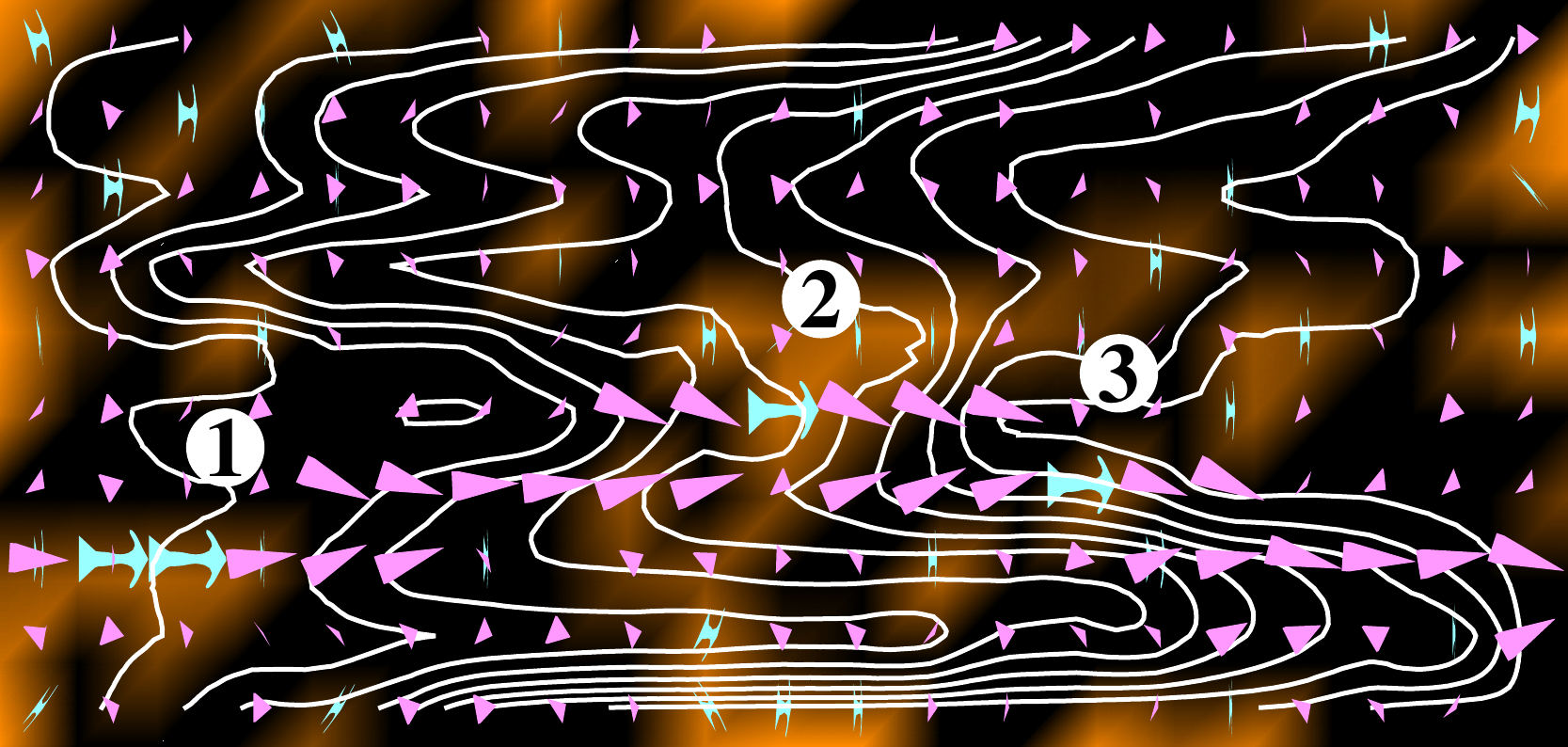}}\\
  (vii) $B=0.09hc/ea^{2}$&(viii) $B=0.10hc/ea^{2}$\\
  \resizebox{0.35\linewidth}{!}{\includegraphics{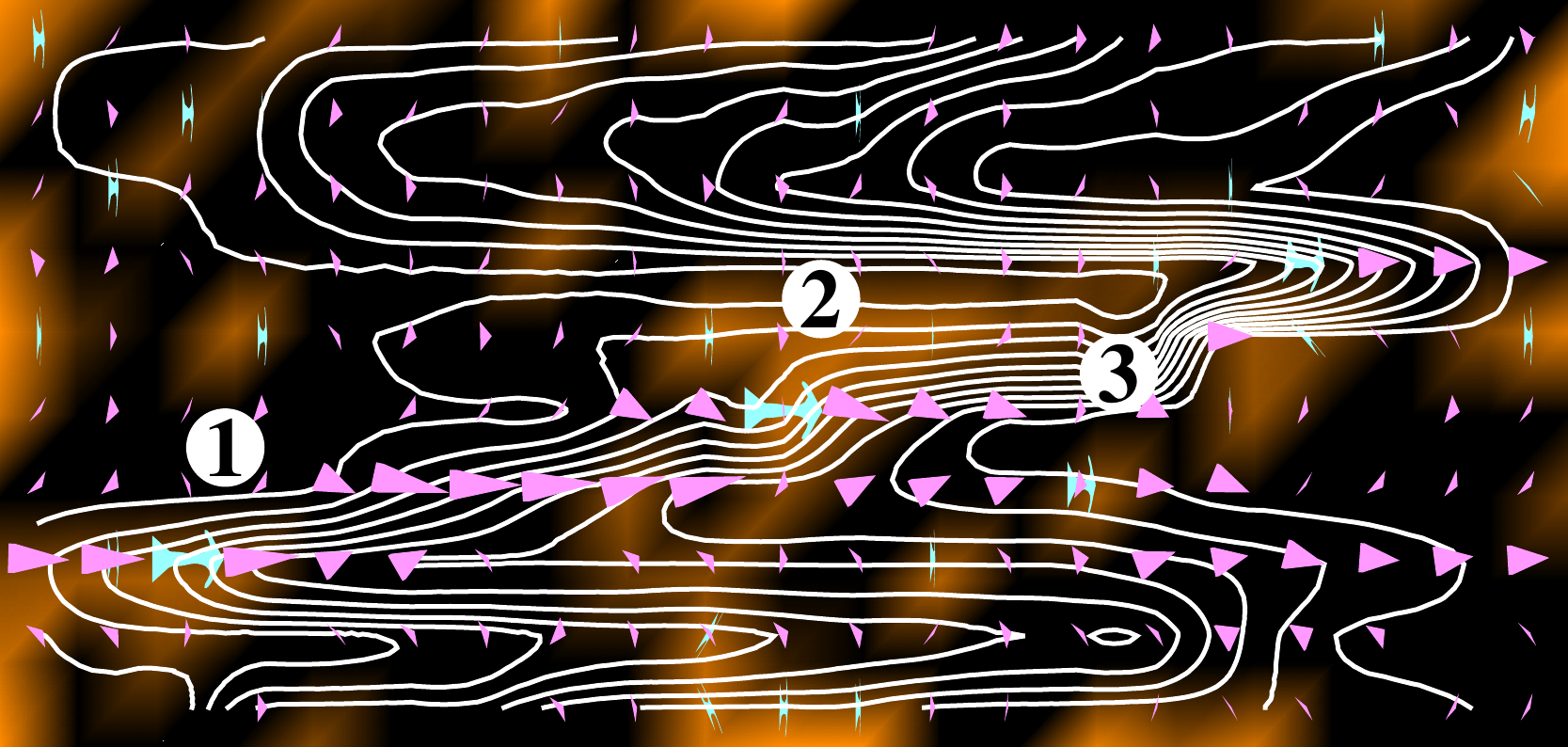}}&
  \resizebox{0.35\linewidth}{!}{\includegraphics{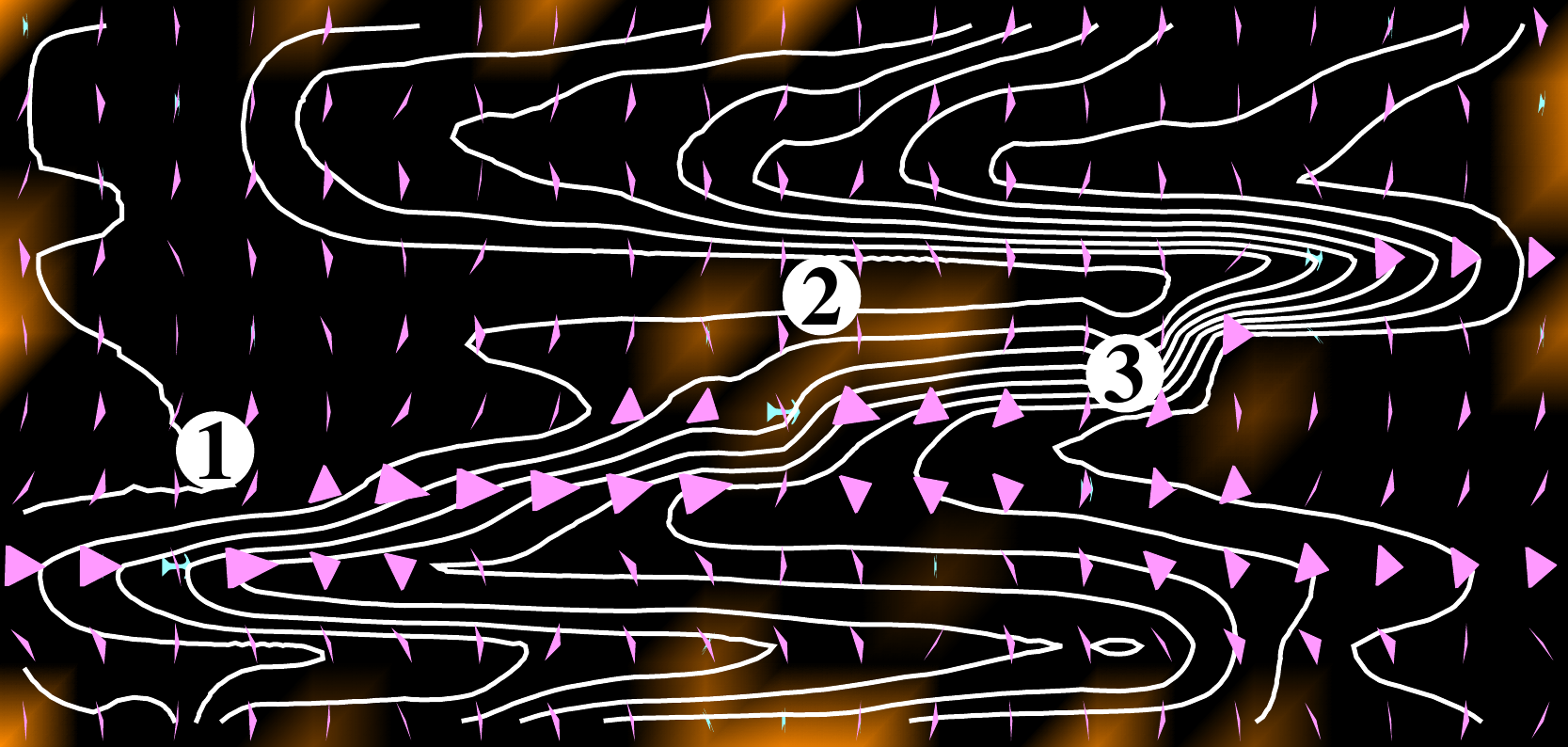}}\\
  (ix) $B=0.11hc/ea^{2}$&(x) $B=0.12hc/ea^{2}$\\
  \resizebox{0.35\linewidth}{!}{\includegraphics{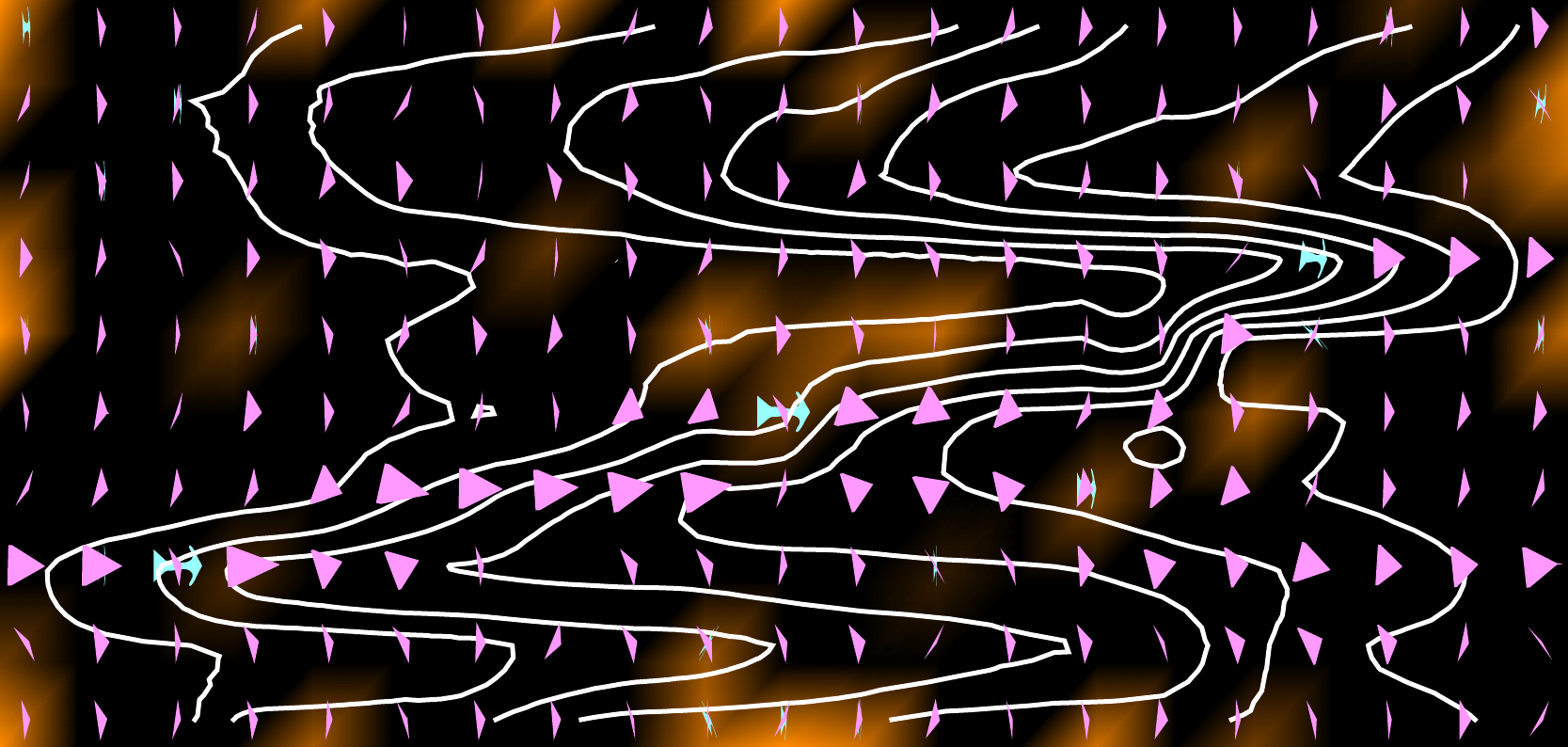}}&
  \resizebox{0.35\linewidth}{!}{\includegraphics{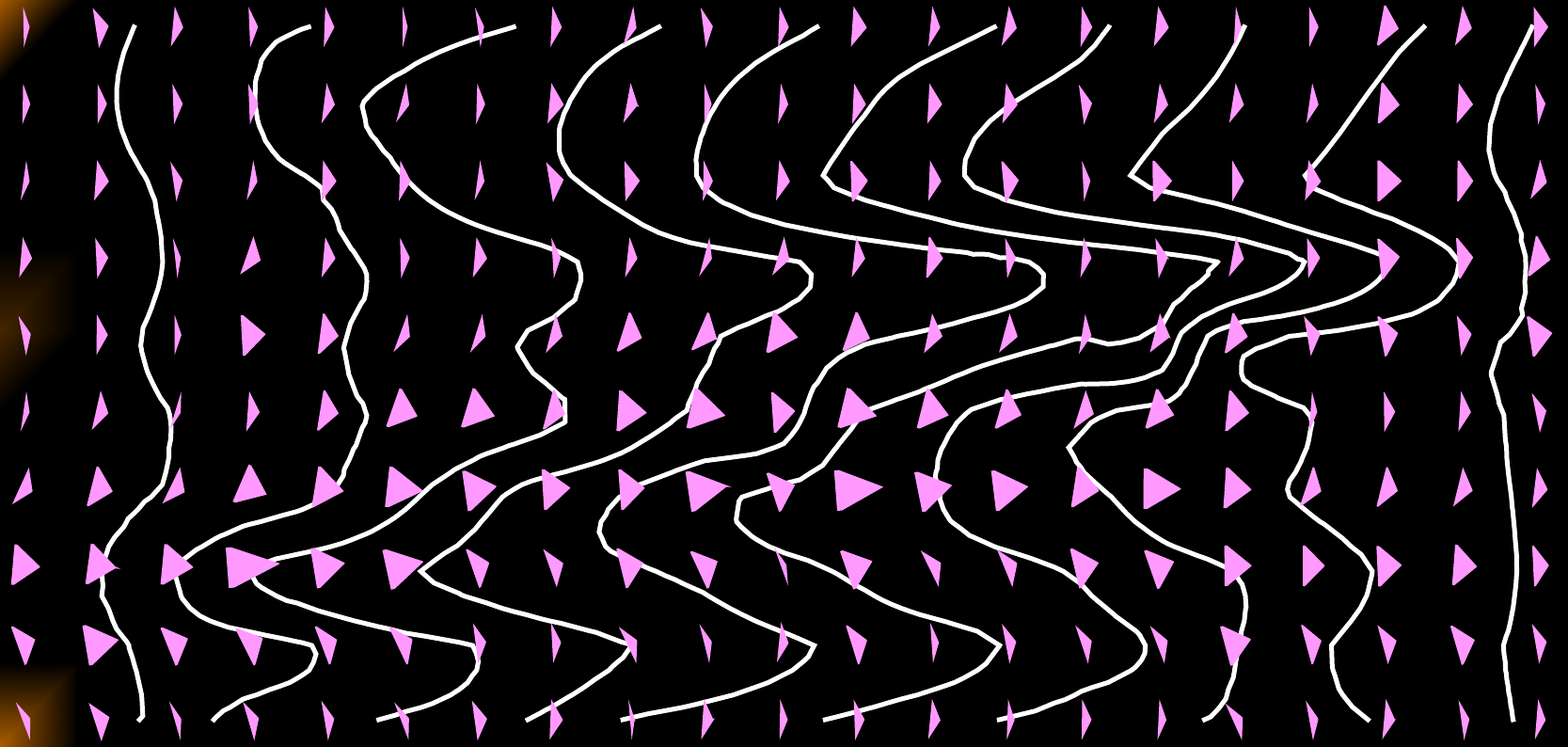}}\\
  (xi) $B=0.15hc/ea^{2}$&Key\\
  \resizebox{0.35\linewidth}{!}{\includegraphics{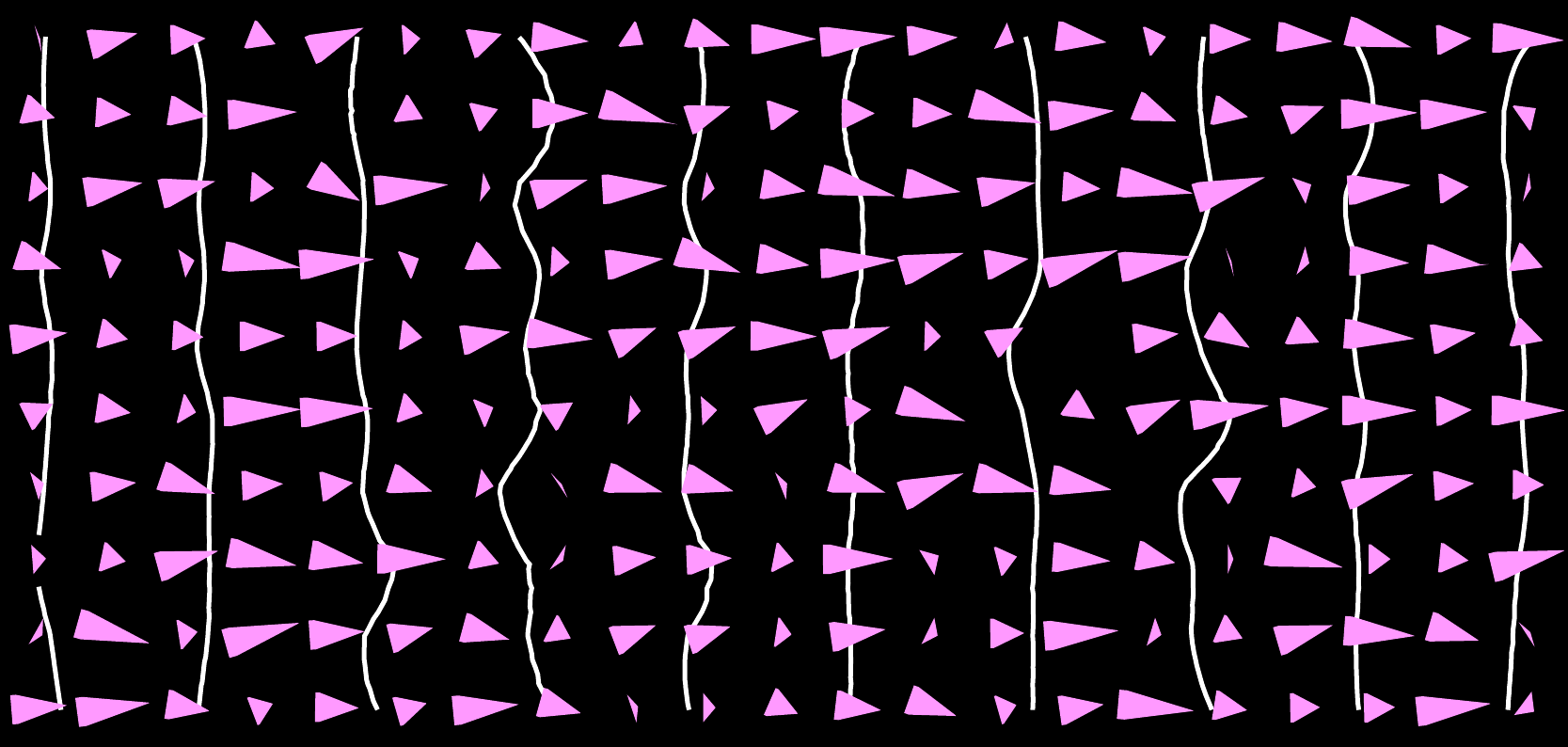}}&
  \resizebox{0.25\linewidth}{!}{\includegraphics{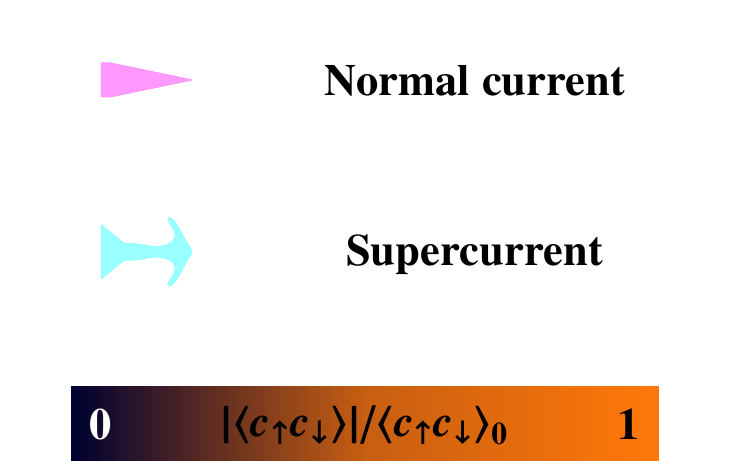}}
 \end{tabular}
 \caption{Current maps at eleven values of the magnetic field shown
 in \figref{fig:Resistance}. The net flow of current is from
 left to right. As illustrated in the key, the cyan darts show supercurrent and
 magenta pointers normal current direction and magnitude; their
 magnitude is renormalized in each map according to the maximum local
 current flow. The white lines show equi-chemical potential lines
 across the sample (ten lines correspond to the full potential between
 the two leads), and background shading depicts the magnitude of the local
 superconducting order parameter ranging from
 strong (orange) to zero (black).}\label{fig:Maps}
\end{figure*}

In this supplement we present a comprehensive series of current maps to
fully expose the microscopic mechanism driving the breakdown of the
superconductor (SC) with applied magnetic field shown in
\figref{fig:Resistance}.  The maps shown in \figref{fig:Maps} cover the
entire scope of the SC-insulator transition: maps (i - ii) capture the
slight rise of resistance as the superconducting order parameter is reduced
but the superconductor remains phase coherent. Maps (iii - vii) outline the
loss of phase coherence and formation of weak links through the sample,
which drives a dramatic increase in the sample resistance. Finally, in maps
(viii - xi) we witness the last remnants of superconductivity being
destroyed and the system entering the normal phase with the resistance
plateauing out.

We first study the maps (i - ii) that show the current flow when the
superconductor is still phase coherent and the resistance is low. The first
map (i), corresponds to $B=0.01$, below the critical field $B_{\text{C}}$
(the unit of the magnetic field is the electron quantum flux, $hc/e$, per
lattice unit area $a^2$). The supercurrent flow is uniform and no voltage is
dropped across the sample, but instead the potential is dropped across the
lead-superconductor junction.  As the field increases to $B=0.04$, (map
(ii)), vortices penetrate the sample at points of strong disorder. The
supercurrent is no longer uniform as it encircles the vortices, but the
sample remains phase coherent and still no potential is dropped across the
sample.

Above the critical field $B_{\text{C}}$, the system loses long-range SC
coherence and the current changes its nature between SC and normal as it
crosses the system.  As the field increases to $B=0.05$ (map (iii)) we see
the first signs of decay of the order parameter and the current flow becomes
normal. With the loss of phase coherence across the system, for the first
time potential is dropped across the sample, though around half of
the total potential difference is still dropped across the
lead-superconductor barriers. Around this field the temperature dependence of
the resistance changes sign from superconducting to insulating.  As the
field increases to $B=0.06$, (map (iv)), normal regions start to form in
the sample, and the potential is now dropped mainly across these weak links.
As the field increases further to $B=0.07$, (map (v)), the disorder
induces specific channels of transport -- the current map in
\figref{fig:Maps} reveals that the current starts as SC, changes into normal
current around point (1) and then splits towards points (2) and (3) where it
reverts to Cooper pairs, only to change back into normal current as it
connects to the right lead. At this magnetic field $B_{\text{M}}$, the
maximum in the resistance, there are approximately equal contributions to
the current from electrons and from Cooper pairs. On increasing the field to
$B=0.08$, (map (iv)), the SC areas (2) and (3) shrink and become the main
source of voltage drop and resistance in the sample. A larger field
$B=0.09$, (map (v)), suppresses the SC correlations, making area (3)
normal, lowering the resistance of that weak link, while at the same time,
the SC area (1) shrinks, giving rise to a higher local resistance. Overall
the resistance of the sample decreases due to the disappearance of some of
the SC island weak links.

Finally, above $B=0.10$ (maps (viii) and (ix)) the SC islands shrink to
such extent that the current flows predominantly around them, through the
normal areas. With a further increase in the magnetic field to $B=0.12$
(map (x)) only normal current flows through the channels that avoid the
residual weak pairing.  However, a further increase in magnetic field to
$B=0.15$, (map (vi)) suppresses superconductivity almost completely,
giving rise to a uniform drop of potential across the sample.

\end{document}